**Evaluation and Refinement of the novel predictive electrolyte model COSMO-RS-ES based on solid-liquid equilibria of salts and Gibbs free Energies of Transfer of Ions**

Simon Müller[a]*, Andrés González de Castilla[a], Christoph Taeschler[b], Andreas Klein[b], Irina Smirnova[a]

* corresponding author (simon.mueller@tuhh.de)

[a] *Institute of Thermal Separation Processes, Hamburg University of Technology, Hamburg, Germany*

[b] *Lonza AG, Visp, Switzerland*

Abstract

*The new predictive electrolyte model COSMO-RS-ES is evaluated and refined for the calculation of solubilities of salts in mixed solvent systems. It is demonstrated that the model is capable of predicting solid-liquid equilibria at 25 °C for ammonium and alkali metal salts quite accurately in a wide variety of solvent mixtures. Furthermore, through the introduction of Gibbs free energies of transfer of single ions it is shown that the model performance can be improved even further. This new data type also allows for an ion-specific way of evaluating the model for the first time. For some systems when calculating the solubility, larger deviations are observed, but for the vast majority of systems the model delivers good predictions. This shows that COSMO-RS-ES is a valuable tool for calculation of phase equilibria in electrolyte systems especially when the scarcity of data impede the application of models that require a higher number of parameters.*

## 1. Introduction

Electrolyte systems are of high importance for various processes in the industry: they are especially relevant for a number of new technologies and the development of efficient separation processes. Therefore, the knowledge of reliable thermodynamic modeling methods of the systems is vital. Whereas the availability of predictive thermodynamic models may help to greatly reduce the experimental efforts, the presence of ionic species poses also challenges to the modeling of these systems.

The calculation of phase equilibria in electrolyte systems has been achieved quite successfully by models that employ binary interaction parameters[1–4], or group contribution based models[5,6]. From the models in literature especially the equations of state can be applied predictively without binary interaction parameters[7,8]. With an ion-based description the number of parameters can be reduced even further as has been published for eCPA[9,10] and ePC-SAFT[11]. Of the group of contribution models AIOMFAC[6] shows capabilities of calculating a large amount of mixed-solvent temperature dependent systems. However, in most cases, the number of parameters needed to represent these systems with higher accuracy tends to increase rapidly when trying to modify conventional thermodynamic models to be able to describe salts in mixed-solvent systems.

An effective model to predict phase equilibria without prior knowledge of the properties of different systems is the $g^E$-model COSMO-RS[12–14]. It relies on statistical thermodynamics and on a general description of the interactions between the molecules. The model was originally developed for neutral molecules. And while it has been shown to be applicable to larger ions (e.g. ionic liquids[15–18], DES[19–21]) very successfully, refinements are necessary to describe electrolyte systems with highly charged ions[22] (e.g. inorganic salts).

There have been two different approaches to model electrolyte systems within the COSMO-RS framework, whereas both of them describe the thermodynamics of close interactions with COSMO-RS and the interactions of ions over longer distance by means of the Pitzer-Debye-Hückel model.

The first approach includes explicit solvent molecules to describe the hydration of the ions and has been employed to calculate pKa values[23], mean ionic activity coefficients (MIAC)[24], hydration energies of ions[25] and vapor-liquid-equilibria (VLE)[26] in aqueous systems. This modelling approach however is difficult to extend to multi-solvent systems due to the inclusion of explicit solvent molecules onto the ions; hence, it has been applied mainly to aqueous systems. Additionally, the exact number of solvent molecules to include into the hydration shell and its conformers in solution is not necessarily known beforehand.

The second approach is based on modified interaction energy equations. This approach has been applied to the calculation of MIAC[26–28] and VLE[27]. This methodology is therefore more empiric than the first one but has no inherent problems to be adapted to multi-solvent systems.

A recent development on predictive electrolyte models using the second approach is COSMO-RS-ES[29]. The model can describe a wide variety of MIAC and LLE systems and is even capable of predicting MIAC, LLE and VLE for systems that were not part of the training set. This shows, that by including LLE data into the training set the activities of the salts can be described much more accurately in mixed-solvent systems. Many models do not include this ternary data, relying on binary data with one solvent and one salt to keep the model as predictive as possible. However, to better describe the salt-concentration dependent transfer of the ions from one phase to another including accurate thermodynamic data such as LLE data is very valuable for the model development.

Based on the good results for the systems already tested, the question arises if the model can be applied to systems that are even more concentrated. In this work, the predictive power of the model COSMO-RS-ES shall be evaluated to calculate solid-liquid equilibrium for fully dissociated electrolytes in mixtures of solvents at 25 °C. Because of the high electrolyte concentration in these systems and the complex interactions possible, these systems pose high demands for any model. Secondly, the inclusion of Gibbs free energies of transfer of ions into the training set is tested to assess the improvements of the performance of the model based on the this new data type. As far as we know, this is an unprecedented way of evaluating an electrolyte $g^E$-model. This new data type allows getting insights of specific ion-solvent interactions which are very valuable to the further development of the model.

## 2. Theory

### 2.1. General relations

Similarly to other electrolyte $g^E$-models the excess Gibbs free energy contribution of COSMO-RS-ES is split into the following terms:

$$\ln(\gamma_i) = \ln\left(\gamma_i^{SR,COSMO-RS}\right) + \ln(\gamma_i^{LR,PDH}) \tag{1}$$

The underlying assumption is that for completely dissociating electrolytes the contribution to the non-ideality of the system can be calculated by using two complementary models. COSMO-RS to describe the so-called short-range (SR) interactions in the immediate vicinity of the species and Pitzer-Debye-Hückel to describe the long-range (LR) coulomb interactions. Here a short summary of both models is given, a more detailed explanation of these might be found in the original publications[29–31].

### 2.2. COSMO-RS

COSMO-RS (COnductor like Screening MOdel for Realistic Solvation)[12,13] is a model that uses the molecular cavities and surface charges calculated with quantum chemistry and the COSMO[32] solvation model in the ideal conductor. The charge density of these surface segments and their surface areas are taken from the COSMO calculations. The model is based on the solution of the statistical thermodynamics of pairwise, independently interacting surface segments. This way complicated and expensive calculations of the interactions between the 3D cavities of the molecules are avoided. To achieve the pairing of the segments first the ideal conductor needs to be removed. However, by this removal, the segments are not ideally screened anymore and a residual charge is left. To describe the interaction energy of this residual charge commonly hydrogen bonding, Van-der-Waals and electrostatic misfit terms are considered.

### 2.3. Pitzer-Debye-Hückel

This model is based on the Debye-Hückel theory that describes the electrostatic interactions between ions in a solution. The extension of the Debye-Hückel theory by Pitzer provides an expression to describe the influence of the long-range interactions between ions on the Gibbs energy. In this model the assumption is made that all ions can be considered as point charges interacting over a distance larger than the closest approach parameter. The solvents are approximated by a dielectric continuum only characterized by its dielectric constant. This model was derived based on the radial distribution function which results from the Debye-Hückel theory as follows[31]:

$$\ln\left(\gamma_i^{LR,PDH}\right) = -\sqrt{\frac{1000}{M_s}}\,\mathrm{A}_\phi\left\{\frac{2z_i^2}{a}\ln\left(1+aI^{1/2}\right)+\frac{z_i^2 I^{1/2}-2I^{3/2}}{1+aI^{1/2}}\right\} \qquad (2)$$

With the Debye-Hückel parameter $\mathrm{A}_\phi$:

$$\mathrm{A}_\phi = \frac{1}{3}\cdot\left(\frac{2\pi N_A \rho_s}{1000}\right)^{1/2}\cdot\left(\frac{e^2}{4\pi\varepsilon_0\varepsilon_s k_B T}\right)^{3/2} \qquad (3)$$

Where $N_A$ is the Avogadro constant, $\rho_s$ the solvent density, $e$ the elementary charge, $\varepsilon_0$ the vacuum permittivity, $\varepsilon_s$ the dielectric constant of the solvent, $k_b$ the Boltzmann constant, $T$ the temperature, $a$ the closest approach parameter and $I$ the mole fraction based ionic strength.

### 2.4. Liquid-liquid-equilibria

As for any neutral species, for electrolytes the following equilibrium equation holds true for a liquid system dividing into an organic phase $O$ and a salt-rich phase $S$:

$$x_i^O \gamma_i^O = x_i^S \gamma_i^S \qquad (4)$$

Where $x_i$ is the mole fraction and $\gamma_i$ the activity coefficient of component $i$ in the respective phase. In the case of a fully dissociated electrolyte the MIAC would be employed.

By introducing the partition ratio $K_i^{OS}$ it might be reformulated like so:

$$K_i^{OS} = \frac{x_i^O}{x_i^S} = \frac{\gamma_i^S}{\gamma_i^O} \tag{5}$$

By iteratively varying the concentration in both phases, it is possible to calculate the LLE by the equilibrium condition expressed in equation (5).

### 2.5. Solid-liquid-equilibria

For a strong electrolyte dissociating as following:

$$M_{\nu_{cat}}X_{an} \leftrightarrows \nu_{cat}M^{z+} + \nu_{an}M^{z-} \tag{6}$$

The chemical potential can be written as:

$$\mu_{salt(solid)} = \nu_{cat}\mu_{cat} + \nu_{an}\mu_{an} \tag{7}$$

Where $\nu$ represent the respective stoichiometric coefficient of each ion. By introducing the definition of the chemical potential $\mu = \mu^+ + \ln(x\gamma^+)$ one gets:

$$\mu_{salt(solid)} = \nu_{cat}[\mu_{cat}^+ + \mathrm{RT}\ \ln(x_{cat}\gamma_{cat}^+)] + \nu_{an}\ [\mu_{an}^+ + \mathrm{RT}\ \ln(x_{an}\gamma_{an}^+)] \tag{8}$$

After refactoring and introducing the solubility product $K_{SP}$, the mean ionic activity coefficient $\gamma_\pm$ and the mean ionic mole fraction $x_\pm$ it can be expressed as:

$$\frac{\mu_{salt(solid)} - \nu_{cat}\mu_{cat}^+ - \nu_{an}\mu_{an}^+ - n\mu_{H2O}^+}{RT} = (\nu_{cat} + \nu_{an})\ \ln\left(x_\pm\gamma_\pm^+\right) = \ln K_{SP}^+ \tag{9}$$

From equation (9) it is possible to see that the solubility product is a function of the reference state and the temperature. This means that if the reference state and the temperature are kept constant, solubility product of a given salt will have the same value in different solvents and mixed solvents.

If the experimental solubility $x_{\pm,ref}$ is known and infinite dilution in a reference solvent (+) is chosen as the reference state, then by calculating the MIAC at the solubility in that solvent $\gamma_{\pm}^{+}(x_{\pm,ref})$ the solubility product can be calculated as:

$$(\nu_{cat} + \nu_{an}) \ln\left(x_{\pm,ref}\gamma_{\pm}^{+}(x_{\pm,ref})\right) = \ln K_{SP}^{+} = (\nu_{cat} + \nu_{an}) \ln\left(x_{\pm,other}\gamma_{\pm}^{+}(x_{\pm,other})\right) \quad (10)$$

Then by iteratively varying the salt concentration in another system $x_{\pm,other}$ and calculating the MIAC $\gamma_{\pm}^{+}(x_{\pm,other})$ it is possible to calculate the solubility in the second solvent when the equilibrium condition is met.

There are also other ways to calculate the solubility product. E.g. if water is chosen as the reference solvent, it is possible to calculate the solubility product by aqueous experimental MIAC values or by Gibbs energies of formation[5]. However, measuring the solubility of a salt in a reference solvent is easy and solubility measurements are abundantly available in literature in comparison to Gibbs energies of formation or MIAC measurements.

For this reason, in this work, infinite dilution in water is chosen as the reference state for all calculations. Since COSMO-RS-ES includes aqueous MIAC values in the training set it can be expected that the solubility products calculated with the model will be similar to experimentally determined values. A similar calculation method of the solubility product is employed by Zuend et al.[33] and is also implemented in the commercial version of COSMO-RS: COSMOthermX.

## 2.6. Gibbs free energies of transfer

The Gibbs free energy of transfer in this case is the difference in chemical potential of the solute at infinite dilution between water and a secondary solvent:

$$\Delta G^0_{t,ion\, w\rightarrow s} = \mu_{ion,s} - \mu_{ion,w} \tag{11}$$

By choosing a common reference state and introducing the definition of the chemical potential this might be rewritten as:

$$\Delta G^0_{t,ion\, w\rightarrow s} = \lim_{x_{ion}\rightarrow 0} RT \cdot \ln\left(\frac{x_{ion}\gamma_{ion,s}}{x_{ion}\gamma_{ion,w}}\right) = RT\ln\left(\frac{\gamma^{\infty}_{ion,s}}{\gamma^{\infty}_{ion,w}}\right) \tag{12}$$

## 3. Computational details of COSMO-RS-ES

**COSMO-RS Calculations**

The interactions between non-ionic species are described with the normal COSMO-RS equations with the electrostatic misfit term using the misfit correlation and the hydrogen bonding energy term[13]. All other equations and their respective parameters involving contacts with ions are shown in Table 1. For a more detailed description of the model please refer to the first publication of COSMO-RS-ES[29].

**Table 1: Interaction energy equations for ionic interactions in the short-range contribution of COSMO-RS-ES.**

| Interaction | Misfit Factor | Ionic Interaction Energy Term |
|---|---|---|
| cation – $H_2O$ | $A_1$ | $E^{ion}_{cat-H_2O} = \frac{a_{eff}}{2} B_1 \sigma_{cat} \max\left(0, \sigma_{H_2O}\right)$ |
| cation – org. mol. | $A_2$ | $E^{ion}_{cat-om} = \frac{a_{eff}}{2} B_2 \sigma_{cat} \max(0, \sigma_{om})$ |
| cation - halide | 0 | $E^{ion}_{cat-hal} = \frac{a_{eff}}{2} B_3 \min\left(0, \sigma_{cat}(1 - D_1 \lvert\sigma_{cat}\rvert^{E_1})\right) \sigma_{hal}$ |
| cation – polyat. an. | 0 | $E^{ion}_{cat-pa} = \frac{a_{eff}}{2} B_4 \min\left(0, \sigma_{cat}(1 - D_1 \lvert\sigma_{cat}\rvert^{E_1})\right) \max\left(0, \sigma_{pa}\right)^{E_2}$ |
| halide - $H_2O$ | $A_3$ | $E^{ion}_{hal-H_2O} = \frac{a_{eff}}{2} B_5 \min\left(0, \sigma_{H_2O}\right) \max(0, \sigma_{hal} - C_1)$ |
| halide – org. mol. | $A_4$ | $E^{ion}_{hal-om} = \frac{a_{eff}}{2} B_6 \min(0, \sigma_{om}) \max(0, \sigma_{hal} - C_2)^{E_3}$ |
| polyat. an. - $H_2O$ | $A_5$ | $E^{ion}_{pa-H_2O} = \frac{a_{eff}}{2} B_7 \min\left(0, \sigma_{H_2O}\right) \max\left(0, \sigma_{pa}\right)$ |
| polyat. an. - org. mol. | $A_6$ | $E^{ion}_{pa-om} = \frac{a_{eff}}{2} B_8 \min(0, \sigma_{om}) \max\left(0, \sigma_{pa} - C_3\right)$ |

### Pitzer-Debye-Hückel Calculations

For mixed solvent systems, the system is treated as a single pseudo-solvent system. A volume fraction mixing rule for the permittivity and a salt-free mole fraction based mixing rule for the density were employed neglecting excess volume effects in both cases. For the closest approach parameter a value of 14.9[31] is fixed.

### Parameterization and evaluation procedure

To parameterize the model a large database of alkali and ammonium salt data was used. This data includes MIAC systems (1190 data points) and LLE systems (1126 data points) at 25 °C. An overview of this data is included in the supporting information.

All parameters of the energy interaction equations and the radii of the cations were optimized simultaneously by means of a Nelder-Mead algorithm.

The following contribution to the objective function for MIAC systems was used in this work:

$$F_{MIAC} = W_{MIAC} \sum_{l} \left( ln\,\gamma_{\pm}^{(m),*,calc} - ln\,\gamma_{\pm}^{(m),*,exp} \right)^2 \qquad (13)$$

The weighting factor $W_{MIAC}$ varied depending on the system type. For alkali halides a value of 40, for all other alkali salts a value of 20 and for the alkaline earth salts a value of 5 was used. These weighting factor were chosen to balance the contribution of the MIAC systems to the objective function in comparison to the LLE and Gibbs free energy of transfer data.

To quantitatively evaluate the models representation of MIAC data the average absolute deviation (AAD) and the average relative deviation (ARD) can be calculated:

$$AAD_{MIAC} = \frac{1}{N_{DP}} \sum \left| ln\left(\gamma_{\pm}^{(m),*,calc}\right) - ln\left(\gamma_{\pm}^{(m),*,exp}\right) \right| \tag{14}$$

$$ARD_{MIAC} = \frac{100}{N_{DP}} \sum \left( \left| 1 - \frac{\gamma_{\pm}^{(m),*,calc}}{\gamma_{\pm}^{(m),*,exp}} \right| \right) \tag{15}$$

For LLE systems, there is a non-iterative way to evaluate the model based on the calculation of the activity coefficients at experimental concentrations according to equation (5):

$$K_i^{OS,calculated} = \frac{\gamma_i^S (x_i^{S,exp})}{\gamma_i^O (x_i^{O,exp})} \tag{16}$$

This way it is possible to compare the experimental partition ratio $K_i^{OS,exp} = \frac{x_i^{O,exp}}{x_i^{O,exp}}$ with the calculated value $K_i^{OS,calculated}$. This evaluation method has two advantages: It is computationally cheaper and independent of the convergence of the iterative algorithm.

Therefore, the following contribution to the objective function was employed:

$$F_{LLE} = \sum \left( ln\left(K_{salt}^{OS,exp}\right) - ln\left(K_{salt}^{OS,calc}\right) \right)^2 \tag{17}$$

While the error of the model was evaluated based on following AAD:

$$AAD_{LLE} = \frac{1}{N_{DP}} \sum \left| ln\left(K_{salt}^{OS,exp}\right) - ln\left(K_{salt}^{OS,calc}\right) \right| \tag{18}$$

For Gibbs free energies of transfer data, the following objective function was used:

$$F_{G-transfer} = \sum \left( \frac{\Delta G^{0}_{t,i\,w\rightarrow s}}{RT} - \ln\left(\frac{\gamma^{\infty}_{i,s}}{\gamma^{\infty}_{i,w}}\right) \right)^2 \tag{19}$$

To characterize the error of the model for the Gibbs free energy of transfer data the following AAD was calculated:

$$AAD_{G-transfer} = \frac{1}{N_{DP}} \sum \left| \frac{\Delta G^{0}_{t,i\,w\rightarrow s}}{RT} - \ln\left(\frac{\gamma^{\infty}_{i,s}}{\gamma^{\infty}_{i,w}}\right) \right| \tag{20}$$

To quickly evaluate the model for its predictive capability for SLE data, it is possible to calculate the expected MIAC that the model should deliver to calculate the correct solubility and compare this value to the one actually calculated.

To calculate the expected MIAC, equation (10) might be rearranged like so:

$$\ln\left(\gamma^{+,expected}_{\pm,other}\right) = \ln\left(x_{\pm,ref}\gamma^{+}_{\pm,ref}\right) - \ln\left(x_{\pm,other}\right) \tag{21}$$

Consequently, the error for the SLE systems was calculated as follows:

$$AAD_{SLE} = \frac{1}{N_{DP}} \sum \left| ln\left(\gamma^{+,expected}_{\pm,other}\right) - ln\left(\gamma^{+,calc}_{\pm,other}\right) \right| \tag{22}$$

In this work whenever the data was used in the parameterization process, the results are considered correlations and whenever the data shown was not part of the training set the results are considered predictions.

**4. New experimental database**

Besides the available database of MIAC and LLE data from the previous work, a database of literature reporting SLE data for electrolyte systems was built concentrating on single salt systems with strong electrolytes containing ammonium and alkali metal salts at 25 °C. The compiled SLE database consists of 224 systems with 785 data points. Only strong electrolytes in aqueous, organic solvents or binary solvent systems were included in the literature search. The references to these systems can be found in the supporting information.

For Gibbs free energies of transfer of ions, the data consists of $Li^+$, $Na^+$, $K^+$, $Rb^+$, $Cs^+$ and $NH_4^+$ cations and $F^-$, $Cl^-$, $Br^-$, $I^-$, $ClO_4^-$, $ClO_3^-$, $SO_4^{-2}$, $SO_3^{-2}$, $SCN^-$ and $NO_3^-$ anions measured at 25 °C. The data for the 222 systems (992 data points) was taken from the following publications: Kalidas et al.[34], Marcus et al.[35], Gulaboski et al[36], Osakai et al.[37], Olaya et al.[38], Marcus[39] and Kihara et al[40].

Moreover, it is possible to calculate further values from experimental data available in literature. The Gibbs free energy of transfer of a salt $\Delta G^{\infty}_{t,salt\,w\rightarrow s}$ is equal to the sum of the Gibbs free energy of transfer of the respective ions. Furthermore, the Gibbs free energy of transfer of a salt might be calculated from experimental values for solubility products in water and another solvent[35]:

$$\Delta G^{0}_{t,salt\,w\rightarrow s} = v_{cat}\Delta G^{0}_{t,cat\,w\rightarrow s} + v_{an}\Delta G^{0}_{t,an\,w\rightarrow s} = \mathrm{RT}\ln\left(\frac{K_{SP,s}}{K_{SP,w}}\right) \tag{1}$$

From equation (1): If the Gibbs free energy of transfer of one ion is known, then with the respective experimental $K_{SP}$ values for the salt, the value for the other ion might be calculated.

In this work, the mole fraction $(x)$ concentration scale is used. However, the available experimental data is commonly presented in molarity scale $(c)$. Thus, the following scale conversion[35] had to be performed for all data points:

$$\Delta G^{0\,(x)}_{t,w\rightarrow s} = \Delta G^{0\,(c)}_{t,w\rightarrow s} - RT\ln\left(\frac{M_w\rho_s}{M_s\rho_w}\right) \tag{2}$$

Where $M_i$ and $\rho_i$ are the molecular mass and the density of the solvent $i$ respectively. In other parts of this work all values presented are in mole fraction scale.

All data compiled was measured at standard conditions (T = 25°C, P = 1 atm).

Results

**First evaluation of COSMO-RS-ES for calculation of salt solubilities**

As a first step the model COSMO-RS-ES was evaluated with the published parameters to assess its predictive power on the new set of solubility data at 25 °C, which are not included into the original training set. Figure 1 A and B show a parity plot comparing the predicted activity coefficient to the expected activity coefficient to calculate the correct solubility calculated according to equation (21).

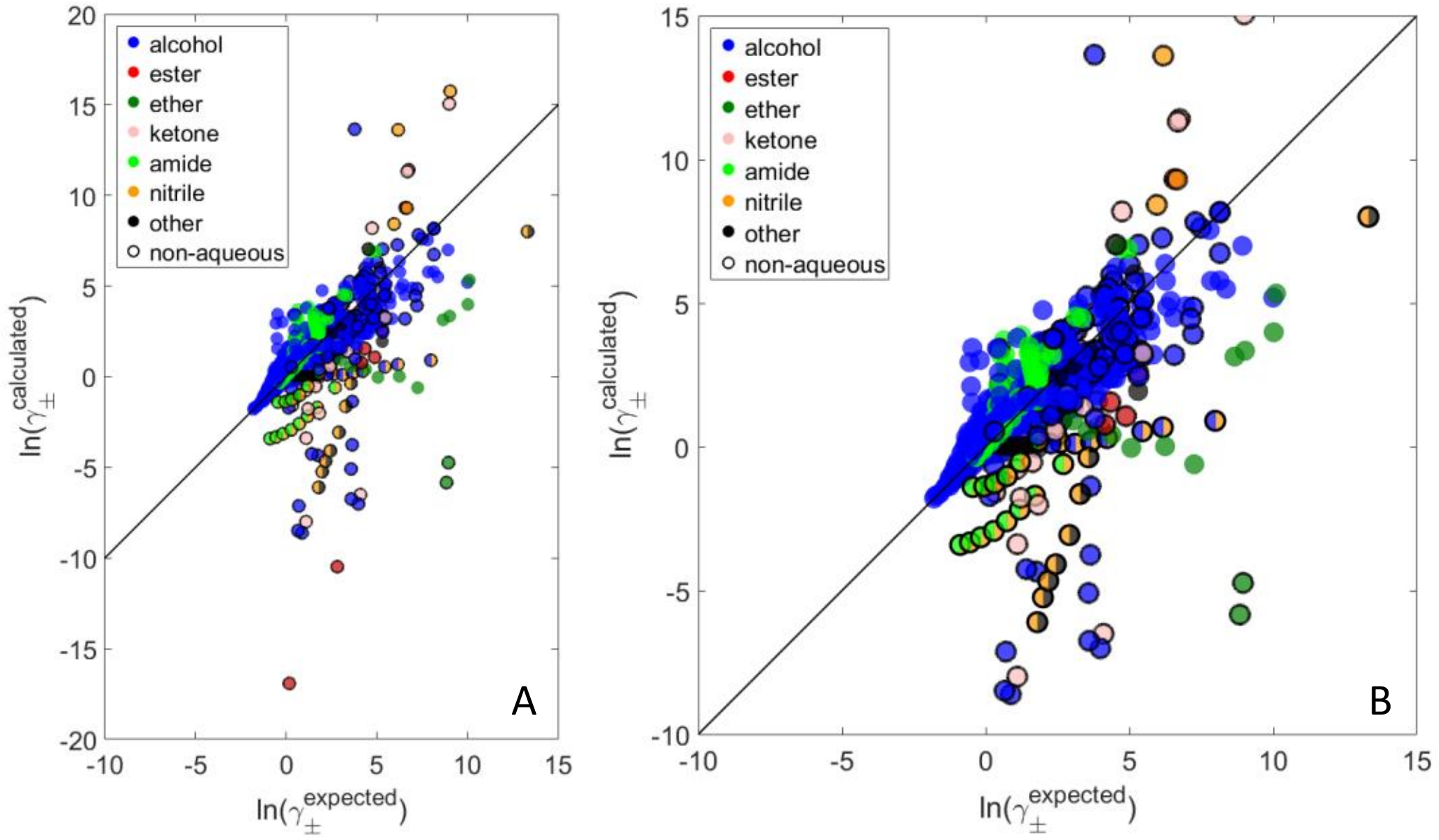

**Figure 1 A: Predicted activity coefficients vs. expected activity coefficients at 25 °C. Calculated with published parameters from Gerlach et al.[29] Solvents are grouped by functional group. If two functional groups were present (e.g. 2-methoxyethanol) the plot shows the system as the non-alcoholic group. Figure 2 B shows an enlarged version of Figure 3 A excluding the largest outliers.**

Over all the error calculated according to the equation (22) is $AAD_{SLE} = 1.23$. It can be observed that in comparison to other solvents the model reproduces the trends for systems with aqueous alcohol systems very well. The training data set of the original publication includes mainly aqueous alcoholic LLE systems. This shows that the model is capable of predicting the activity coefficients needed for the solubility calculation in such systems. Also, for high permittivity solvents like the amides mixed with water the representation is good although a slight systematic overestimation of the calculation can be seen. For non-aqueous systems, however the model predictions lie far away from the expected value. Especially for systems with solvents of the functional groups ethers, ketones, nitriles and esters the error is large. The largest errors are for the solubilities of $LiClO_4$ in diethyl ether and in ethyl acetate. Common to these systems is that these solvents have an extremely high solubility of the specified salt (around 50 wt.-%) and a low permittivity ($\epsilon$ around 5). These types of systems take the model to the limit of its capabilities due to these extreme conditions. At high concentrations and low permittivities effects like ion-pairing may become very prominent[41,42] but are so far neglected. Including these effects into the modelling approach could enhance the description of these systems. Nevertheless, for the vast majority of systems the model already performs very well, taking into account that no specific binary parameters for the corresponding solvent are needed.

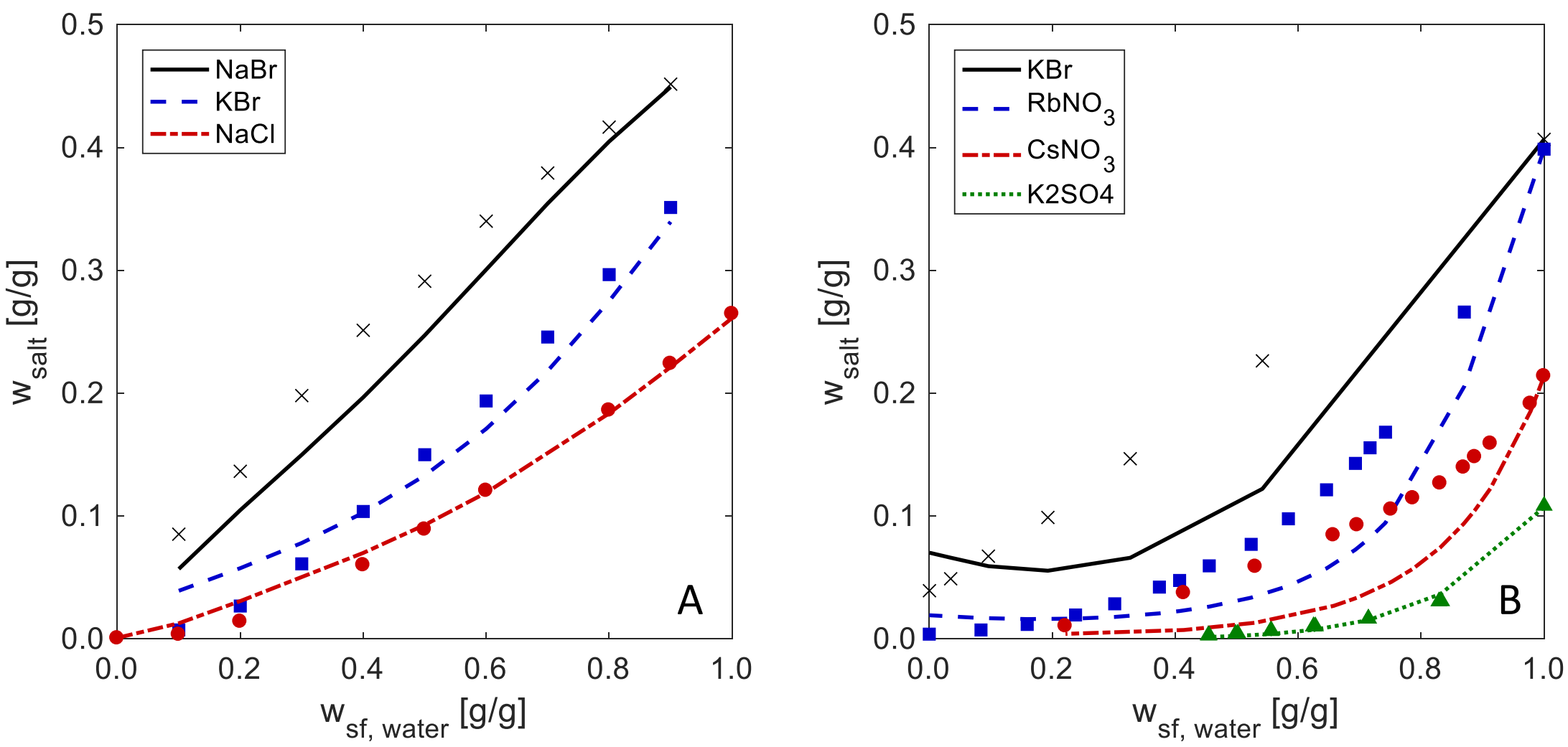

**Figure 4: Solubilities at 25 °C predicted by COSMO-RS-ES plotted vs. the salt-free water concentration. Calculated with published parameters from Gerlach et al.[29] Markers correspond to the experimental values and lines to the calculated values. Figure 4 A shows the following systems: NaBr in water + ethanol[43], KBr in water + ethanol[44], NaCl in water+ ethanol[45]. Figure 4 B shows the following systems: KBr in water + 2-methoxyethanol[46], $RbNO_3$ in water + methanol[47], $CsNO_3$ in water + tert-butanol[48], $K_2SO_4$ in water + 1-propanol[49].**

Figure 4 shows several examples of solubilities calculated with COSMO-RS-ES using the published parameters. The model is capable of quantitatively predicting the salt solubility of sodium bromide, potassium bromide and sodium chloride in mixtures of ethanol and water for alkali halides. Even for systems with more complex salts like rubidium nitrate, cesium nitrate and the doubly charged potassium sulfate in different solvents the model predicts the qualitative trend correctly in all cases and in some cases it quantitatively agrees with the experimental data. For these systems an underestimation of the solubility can be seen the higher the experimental solubility is. This might be due to the overestimation of the activity coefficient of the salt with rising organic solvent content as well as an underestimation of the different preferential solvation of one solvent over the other. A similar effect was already observed in the previous publication[29] where the model predicted the tendency of the salt to prefer the salt-rich phase, which in most cases is the water phase.

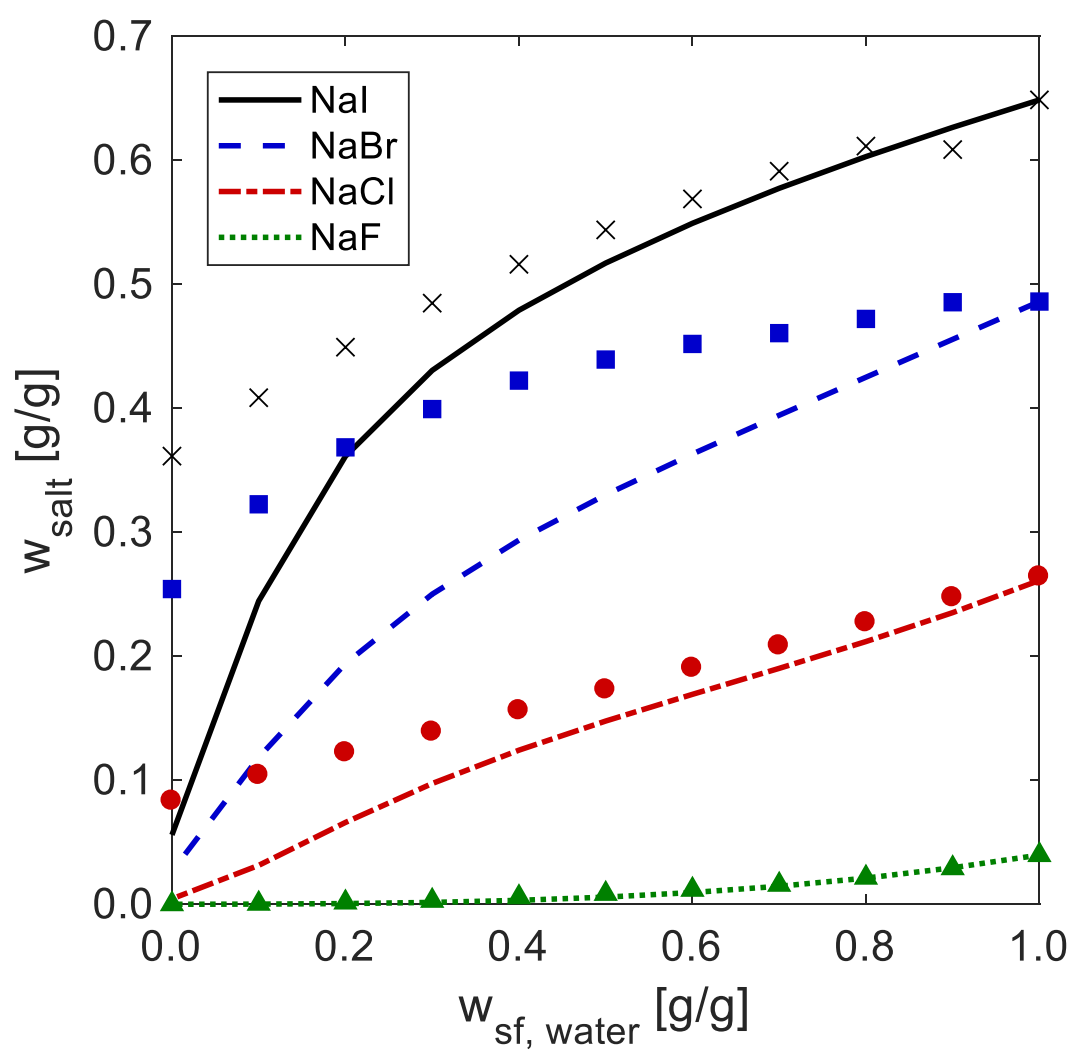

**Figure 5: Solubilities at 25 °C predicted by COSMO-RS-ES using the published parameters vs. the salt-free water concentration. Markers correspond to the experimental values and lines to the calculated values. Systems shown are NaI, NaBr, NaCl, NaF in mixtures of water and formamide[50].**

In the training data set of the previous publication (MIAC and LLE) almost every system included some water. With the new dataset of solubilities tested in this work a lot of systems that do not include water at all can be tested. According to equation (10), the model describes the activity of a salt in what can be a completely non-aqueous system with the reference state infinite dilution in water. Figure 5 shows the solubilities calculated for the sodium halides in mixtures of water and formamide. It can be observed for higher solubilities, that the less water is present in the system, the larger the deviation of the model gets. The solubility of the different salts in pure formamide is reproduced in the correct order, but the values are not quantitatively accurate. This is due to the larger deviation of the calculations for pure organic solvents and could already be observed in Figure 1 on a much larger scale.

### 4.1. Evaluation of COSMO-RS-ES for Gibbs free energies of transfer

In this work, the solubilities are modelled by using water as a reference solvent as described above. Thus, to correctly calculate the solubility in another solvent, the transfer of the salt from water to this solvent needs to be described accordingly. Although this transfer is not at infinite dilution, the Gibbs free energy of transfer of ions poses a way to assess how well a part of this transfer is described.

One contribution that has not been considered so far within the COSMO-RS-ES model which could have an influence on the description of the Gibbs free energies of transfer is the Born term. Some $g^E$-models include this contribution[51–53], also many recent EOS[3,54–56] employ this term to better describe the effect of the permittivity difference of the solvents on the calculation of solvation energies. COSMO-RS itself might be considered as an improved theory in comparison to continuum solvation models[14], which raises the question if a Born term should be used. However, the correct inclusion of this term can be somewhat difficult. Assuming the permittivity is not salt concentration dependent the inclusion of the term is less complicated, but it is known that salts have a large influence on the permittivity of the solution. One could think of using a model based on experimental dielectric decrement data, but as Kontogeorgis et al.[42] and Maribo-Mogensen et al.[57] point out, even the experimental data for the dielectric decrement would have to be evaluated carefully due to the effect of kinetic depolarization during the measurement. Because of these complexities and to test whether the model is capable of describing the systems without a Born term, this contribution was neglected in this work.

Data previously employed to evaluate the model was always salt specific data based on the calculation of the mean ionic activity coefficients. This allowed attributing effects observed to specific salts. However, since this new type of data is ion-specific it allows for the first time to evaluate the model more thoroughly by attributing effects to specific ion groups. This allows for a much more detailed way to refine the model since the energy interaction equations are based on ion groups.

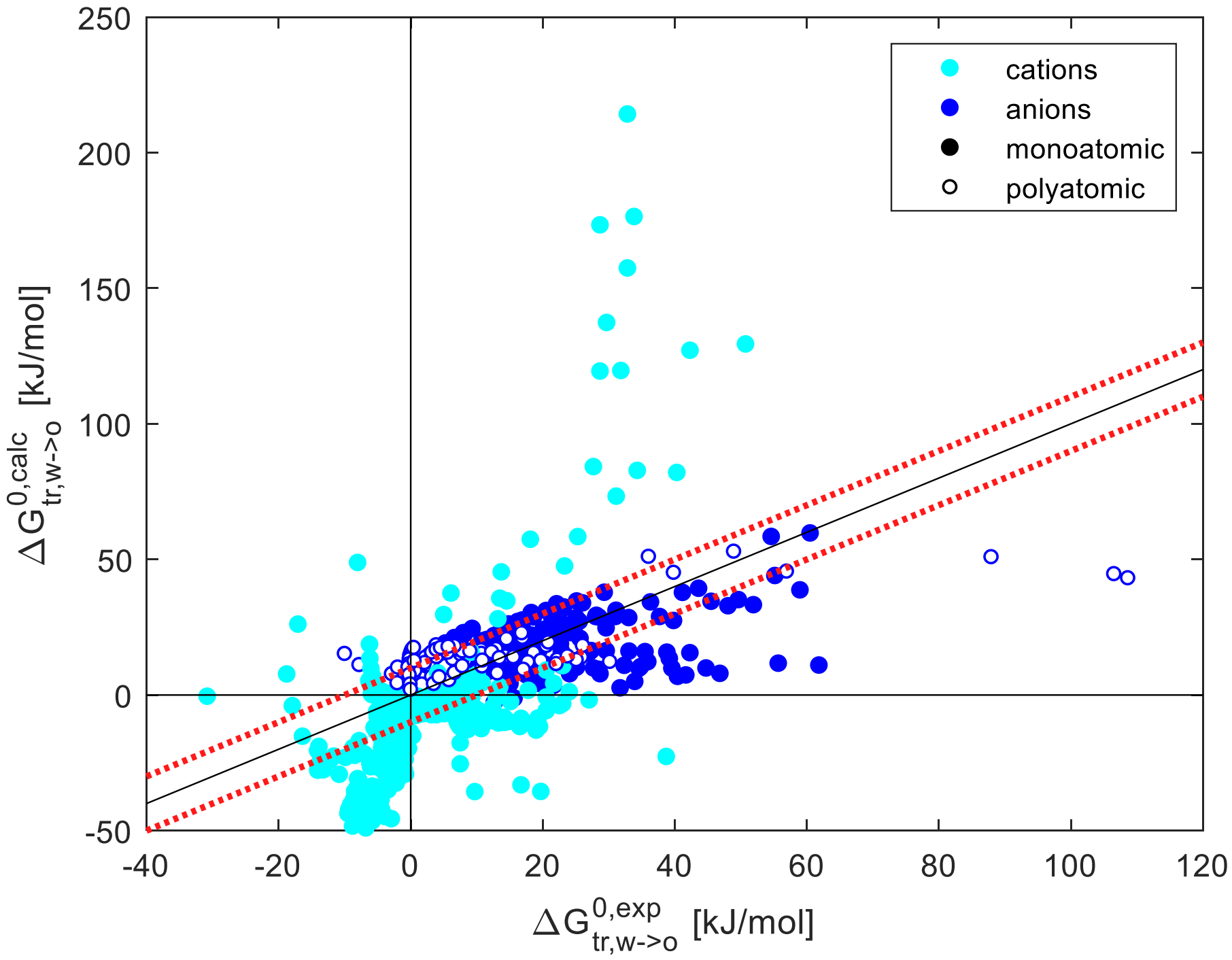

**Figure 6: Gibbs free energies of transfer ions (mole fraction based) at 25 °C predicted with COSMO-RS-ES using the published parameters plotted against the experimental values. Red dotted lines denote a range of ±10 kJ/mol for the maximum estimated experimental error according to Kalidas et al[34]. Filled circles (●) denote monoatomic ions while open circles (○) represent polyatomic ions.**

The evaluation of the predictive capabilities of COSMO-RS-ES to calculate Gibbs free energies of transfer of ions with the published parameters can be seen in Figure 6. According to the different compilations of this datatype used in this work, the accuracy of the experimental values lies between ±3 and ±10 kJ/mol. Although the error of anions is generally estimated to be lower than the one from cations, the maximum error of ±10 kJ/mol was included into the graph to better discuss the results of the model.

Figure 6 shows that the anions are predicted quite accurately ($AAD_{G-transfer}^{anion} = 2.79 \frac{kj}{mol}$) having the majority of the calculated values fall into the experimental uncertainty. Independently of the ion size (monoatomic halide or polyatomic anion), both seem to be represented with similar accuracy. The two largest outliers are for the sulfate anion in dimethylformamide and acetone.

The calculated values for the cations on the other hand, some values are calculate correctly but others show large deviations ($AAD_{G-transfer}^{cation} = 5.86 \frac{kJ}{mol}$) even to almost an order of magnitude. This shows that the model is not describing the interactions of the cations with the organic solvent correctly when compared to water.

Overall the model however is capable of describing some trend with a combined error of $AAD_{G-transfer} = 4.35\,\frac{kJ}{mol}$. All AAD values are lower than the assumed experimental error, but it can be observed that for the cations the large number of outliers might lead to wrong predictions when calculating solubilities.

Because of the extrathermodynamic assumptions needed for splitting the energies into its ionic components, this data is not as reliable as other experimental measurements, but it makes this sort of ion specific analysis important for the possible development of the model.

### 4.2. Inclusion of Gibbs free energies of transfer into the training set

To our best knowledge, this is the first time Gibbs free energies of transfer are used to improve the accuracy of a predictive electrolyte $g^E$-model. Thereby, several parameterization strategies were tested. The best overall results were achieved by using a two step optimization. First, the model was adjusted to the complete dataset including MIAC, LLE and Gibbs free energies of transfer using the published parameter values as starting point. Due to the large relative experimental error in the Gibbs free energy of transfer, we consider the values of the Gibbs free energies of transfer as a guide to qualitatively improve the model, hence, in the second step we adjust the model to MIAC and LLE data only using the parameter values from the first step as starting point. Table 2 shows the resulting parameters of the new parametrization. As expected, the parameters that change the most ($A_2$ and $B_2$) describe the interactions between the cations and the solvents. It was possible to achieve a much better correlation of the Gibbs free energy of transfer data (Figure 7) than with the published parameters by still describing the other systems (MIAC and LLE) without a loss in accuracy.

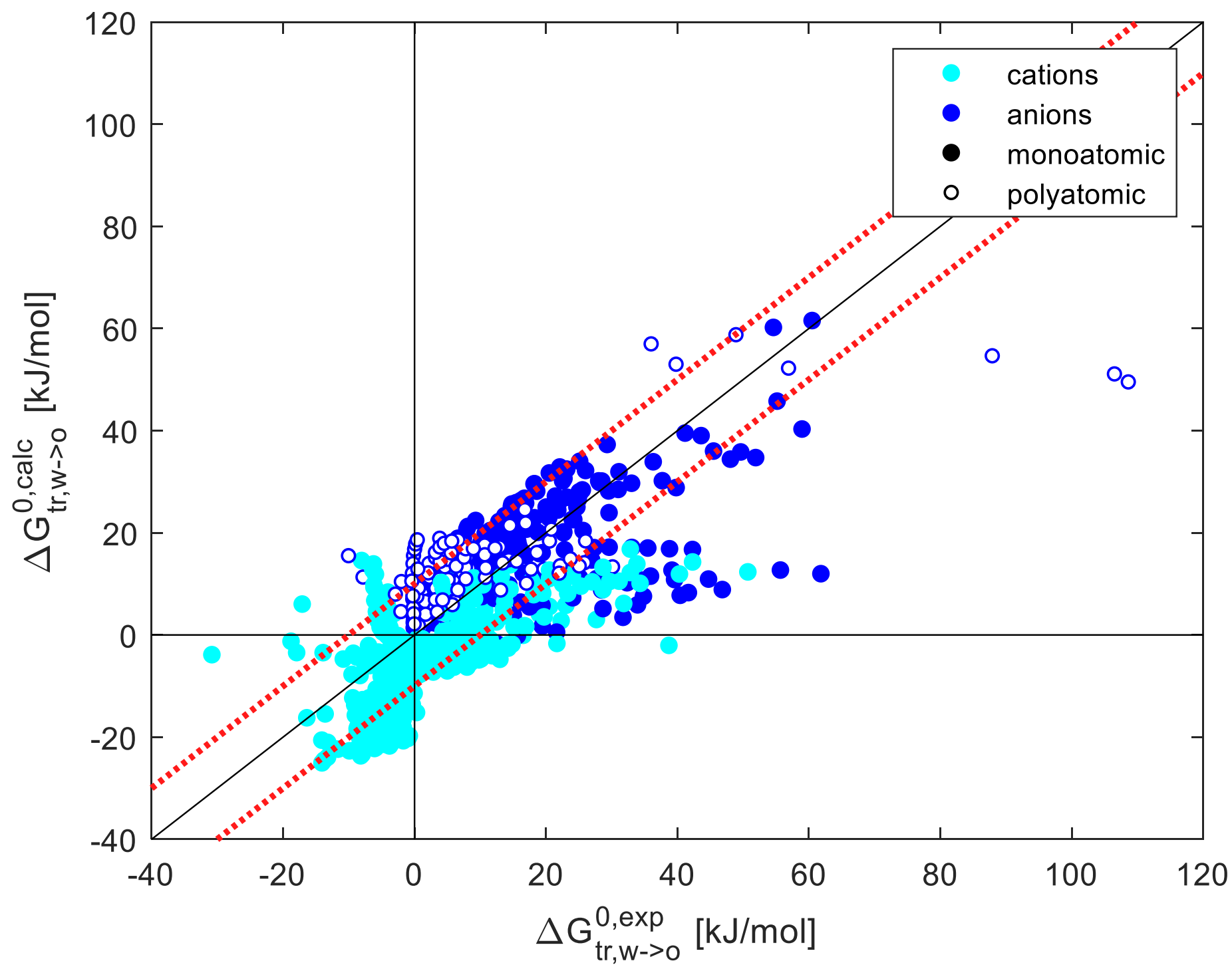

**Figure 7: Gibbs free energies of transfer ions (mole fraction based) at 25 °C correlated with COSMO-RS-ES using the adjusted parameters plotted against the experimental values. Red dotted lines denote a range of ±10 kJ/mol for the maximum estimated experimental error according to Kalidas et al[34].**

| cation radii [Å] | | parameters A,B $\left[\frac{\mathrm{kJ}\ \text{Å}^2}{\mathrm{mol}\ \mathrm{e}^2}\right]$ | | parameters C $\left[\frac{\mathrm{e}}{\text{Å}^2}\right]$ | | parameters D,E $[-]$ | |
|---|---|---|---|---|---|---|---|
| $Li^+$ | 1.752 | $A_1$ | 7178 | $C_1$ | 0.0020 | $D_1$ | 562.7 |
| $Na^+$ | 1.901 | $A_2$ | 3300 | $C_2$ | 0.0107 | $E_1$ | 1.747 |
| $K^+$ | 2.015 | $A_3$ | 2088 | $C_3$ | 0.0321 | $E_2$ | 0.015 |
| $Rb^+$ | 2.076 | $A_4$ | 6374 | | | $E_3$ | 0.332 |
| $Cs^+$ | 2.285 | $A_5$ | 3459 | | | | |
| | | $A_6$ | 2538 | | | | |
| | | $B_1$ | 6639 | | | | |
| | | $B_2$ | 186 | | | | |
| | | $B_3$ | 10327 | | | | |
| | | $B_4$ | 99 | | | | |
| | | $B_5$ | 10613 | | | | |
| | | $B_6$ | 1362 | | | | |
| | | $B_7$ | 4741 | | | | |
| | | $B_8$ | 4397 | | | | |

**Table 2: New parameterization of COSMO-RS-ES including the Gibbs free energies of transfer of ions into the parameterization procedure**

The overall error of the LLE systems ($\boldsymbol{AAD_{LLE} = 0.69}$) and the MIAC systems ($\boldsymbol{ARD_{MIAC} = 4.37}$) remain equal to Gerlach et al.[29] While the error for some LLE systems increases slightly, for other (e.g. ketone) LLE systems the description of the LLE is improved very much by the new parameters. This leads to the overall error staying the same, but to the trends being described better for a wider range of solvent types. Figure 8 shows an example of this improved performance.

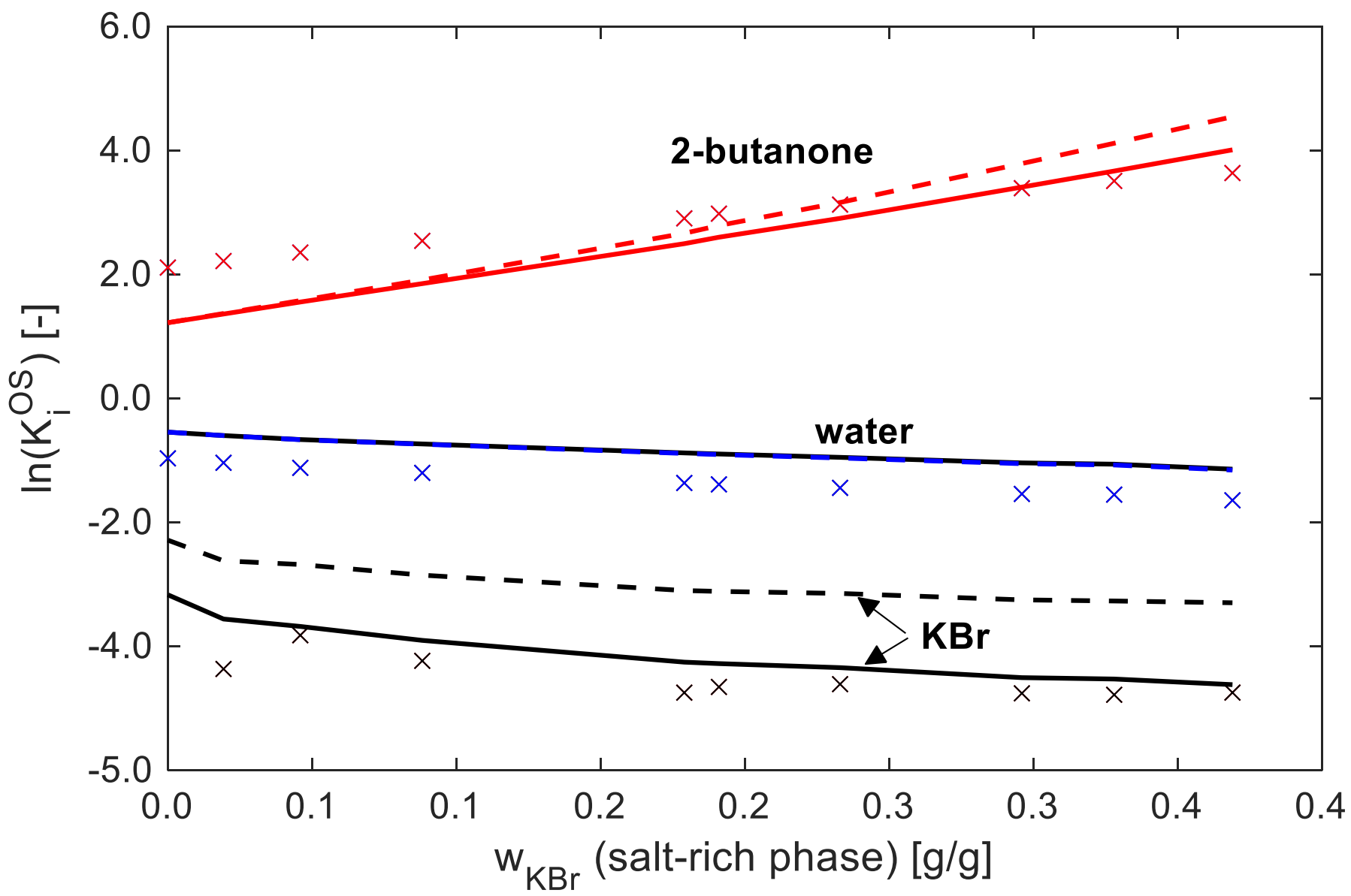

**Figure 8: Partition of components in the LLE system potassium bromide, water and 2-butanone at 25 °C correlated by COSMO-RS-ES according to equation (16) in comparison to the experimental values[58]. The markers correspond to the experimental values, the dashed lines represent calculations with published parameters from Gerlach et al.[29], solid lines represent the calculated values with the new parameters.**

For the solubility calculation however, the improvement is even larger. The average absolute deviation drops to $AAD_{SLE} = 0.98$. As can be seen in Figure 9, especially the non-aqueous systems are predicted in better accordance to the expected value. Although by introducing the Gibbs free energies of transfer to the training dataset only the transfer of a salt at infinite dilution from water to another solvent is expected to be described better, this seems also to have an impact on how well it describes the activity coefficient at finite concentrations.

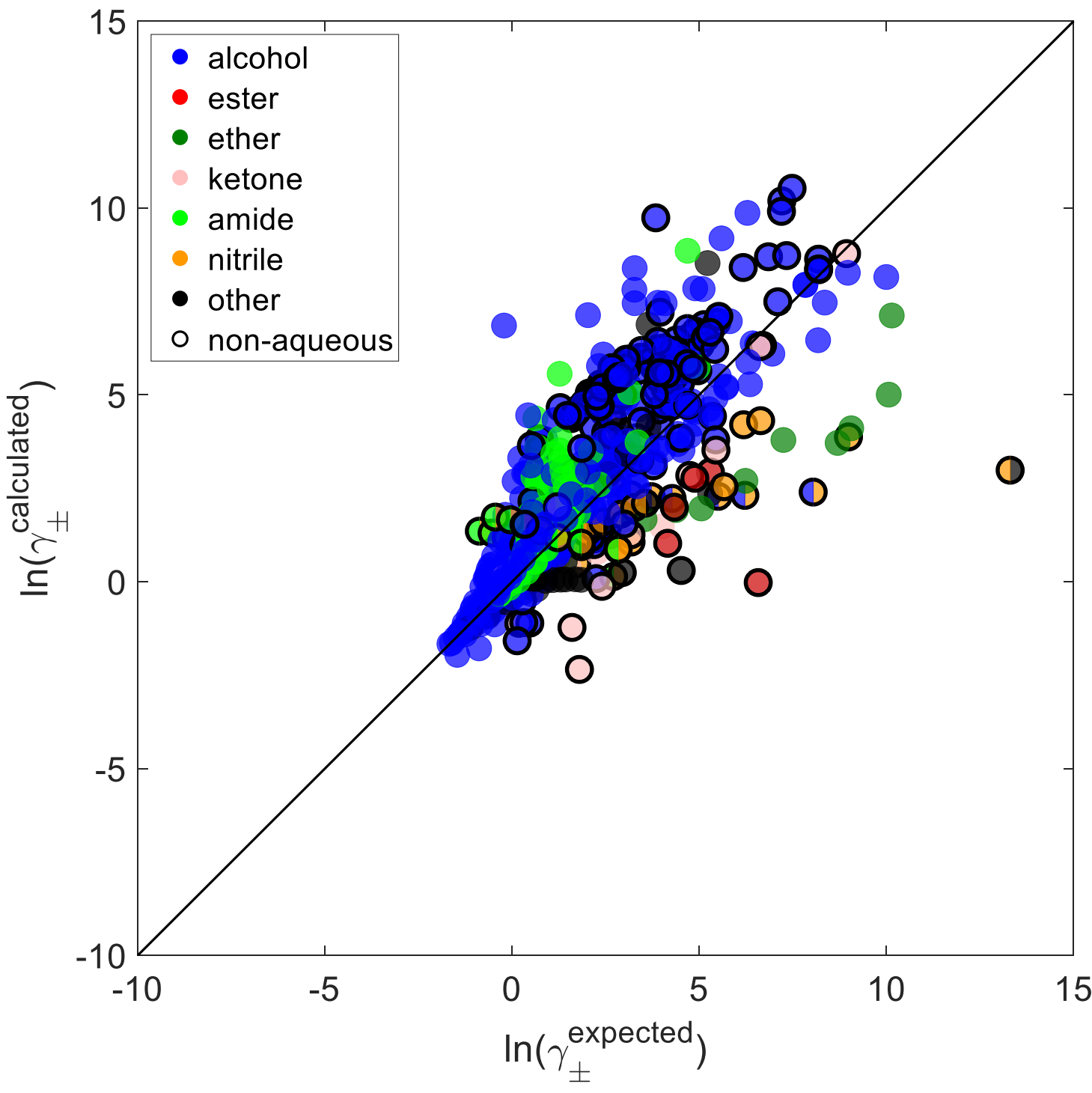

**Figure 9: Predicted activity coefficients vs. expected activity coefficients at 25 °C calculated with new parameters (Table 2). Solvents are grouped by functional group. If two functional groups were present (e.g. 2-methoxyethanol) of which one is an alcoholic group, the plot shows the system colorized as the non-alcoholic group.**

The two outliers of ester systems seen in Figure 1 are still outside of the scale of this figure. There still seems to be an issue with describing the extremely high solubilities of salts in some low permittivity solvents. However, for most systems the new parameters allow a more accurate prediction of the salt solubilities than with the parameters from our previous work. At very high concentrations, the balancing of the short-range term and the long-range term needs to be reviewed as the used long-range term in COSMO-RS-ES was developed for rather lower salt concentrations. Furthermore, at high concentrations because the ion-ion contact becomes more probable, a refined description of these contacts might be necessary. Until now no special equation is being used to treat the different possible interactions between cations and anions with each other.

Figure 10 shows various examples of solubilities predicted with the new parameter set in comparison with the parameters from Gerlach et al.[29] It can be observed that the trends improve for different types of salts with monoatomic and polyatomic anions as well as for different solvents. In Figure 10 A for the systems with potassium bromide and cesium nitrate the model predicts more accurate results for mixtures of water and an organic solvent while for the system with potassium chloride especially at low water concentrations, the model performs better. A similar progress can be seen in Figure 10 B for the system with NaI. Although the solubility of the salt is very high (up to 70 wt.-%) the values are predicted with better accuracy than before.

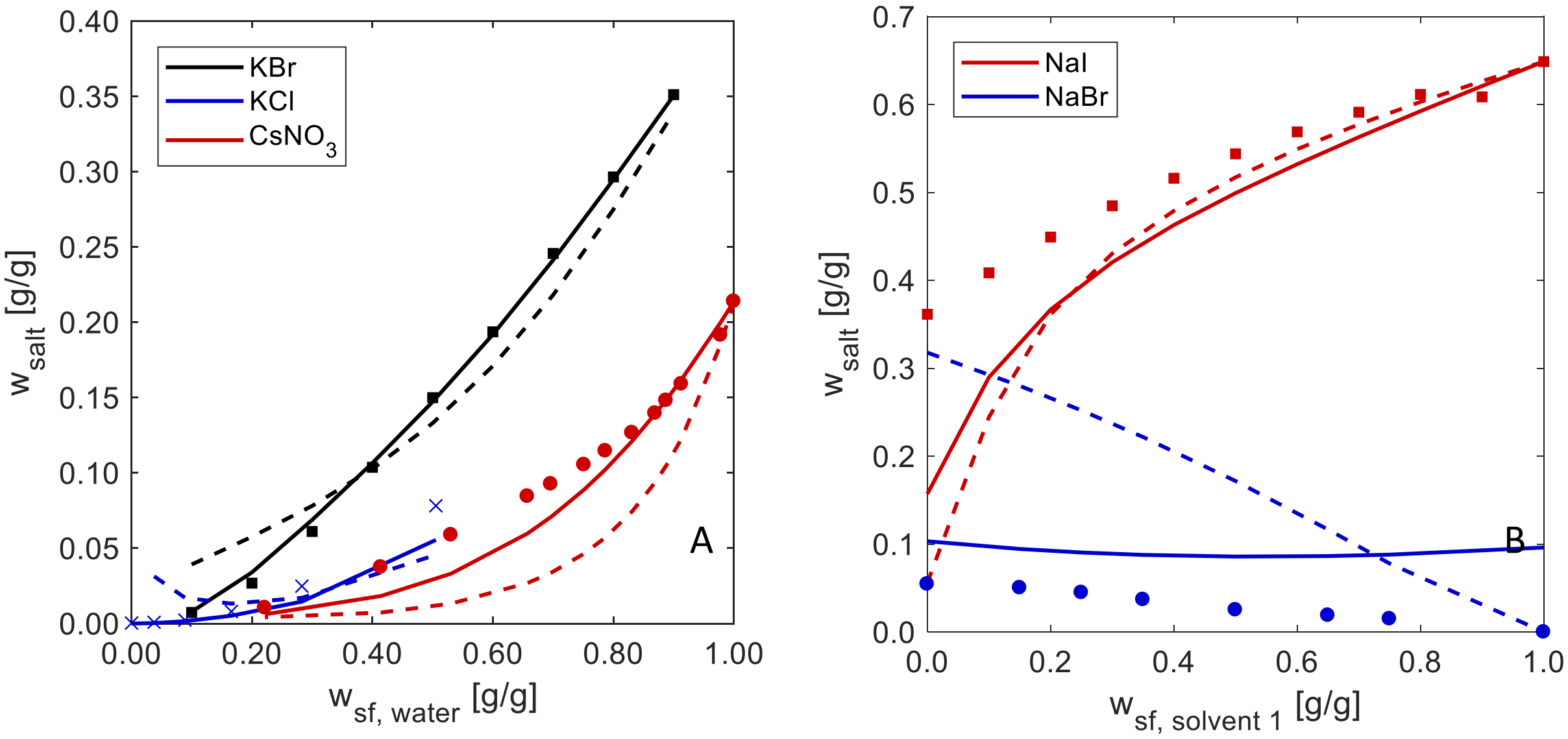

**Figure 10: Solubilities at 25 °C predicted by COSMO-RS-ES vs. the salt-free solvent concentration: (A) - KBr in water + ethanol[44], KCl + water + 2-isopropoxyethanol, $CsNO_3$ in water + tert-butanol[48]; (B) - NaI + water(1) + formamide(2)[50] and NaBr + acetonitrile(1) + DMSO(2)[59]. The markers correspond to the experimental values, the dashed lines represent calculations with published parameters from Gerlach et al.[29], solid lines represent the calculated values with the new parameters.**

The non-aqueous system sodium bromide in mixtures of acetonitrile and DMSO (Figure 10 B) shows mixed improvements. Although the solubilities in DMSO rich mixtures are described better, the error becomes larger at higher concentrations of acetonitrile. For this system, the activity coefficients of the salt are not described accurately, this can already be seen in Figure 9 where the point of this system in pure acetonitrile is the one with the largest error.

Deviations in completely water free systems like those discussed above might be better described in the future by including osmotic coefficients into the training data set to improve the description of the activity of salts in these types of systems. Until now, only data points that include water have been used to train the model and osmotic coefficients give the needed information about the activity of the salt in non-aqueous systems as the salt concentration changes.

In general, although the predictions are not perfect in every case, they show the potential of the model to describe wide varieties of systems with a relatively small set of parameters without relying on binary interaction parameters.

## 5. Conclusions

This work demonstrates the predictive capabilities of the new electrolyte model COSMO-RS-ES for the calculation of salt containing systems up to the solubility limit. The model is based on a modification of the widely used model COSMO-RS. These modifications are the introduction of a set of interaction equations and a combination with the Pitzer Debye-Hückel model to describe the ionic long-range interactions. It was shown that the model can predict the solubilities in a range of systems based on the parameters already published by Gerlach et al.[29] All relevant parameters of the model are published within this paper and previous work by Gerlach et al.[29]

Furthermore, experimental data for Gibbs free energies of transfer of ions were firstly introduced for the model parametrization. With this type of data, it was possible to analyze the model in an ion specific form for the first time. This data showed that the values for the anions are already quite accurate, while for the cations large errors were found.

In the next step the new data type was included into the training set of the model. This lead to an even better representation of this data, especially for the cations, while still describing the MIAC and LLE systems with an overall accuracy equal to the earlier publication. For the LLE systems some ketone systems were calculated more accurately than before.

The prediction of the solubilities was improved in most cases by the addition of the new data. Especially the trends for non-aqueous data were improved. Some deviations were found in the description of completely non-aqueous systems particularly at high solubilities and low permittivities.

Overall COSMO-RS-ES proves a versatile model able to describe electrolyte systems in a predictive manner with a small amount of general parameters, thus becoming particularly useful in cases where no or just scarce experimental data is available which would be necessary to train other models. Therefore, the model provides the possibility to accelerate the development of processes with less common salts for which no parameters are available in other thermodynamic models.